\documentclass[11pt]{article}
\usepackage{amsmath,amssymb,color,cite}
\usepackage{graphicx}
\usepackage{amsfonts}
\usepackage{enumerate}
\usepackage{bbm}
\usepackage{multirow}
\usepackage{nicefrac}
\usepackage[all]{xy}
\usepackage{graphicx}
\usepackage{booktabs}
\usepackage{hyperref}
\usepackage{amsfonts}

\newcommand{\ra}[1]{\renewcommand{\arraystretch}{#1}}
\newcommand{\hoch}[1]{$\, ^{#1}$}
\newcommand{\cN}{{\cal N}}

\makeatletter
\@addtoreset{equation}{section}
\makeatother

\newcommand{\be}{\begin{equation}}
\newcommand{\ee}{\end{equation}}
\newcommand{\bea}{\setlength\arraycolsep{2pt} \begin{eqnarray}}
\newcommand{\eea}{\end{eqnarray}}
\newcommand{\nn}{\nonumber}

\def\ft#1#2{{\textstyle{\frac{\scriptstyle #1}{\scriptstyle #2} } }}

\def\0{{\sst{(0)}}}
\def\1{{\sst{(1)}}}
\def\2{{\sst{(2)}}}
\def\3{{\sst{(3)}}}
\def\4{{\sst{(4)}}}
\def\5{{\sst{(5)}}}
\def\6{{\sst{(6)}}}
\def\7{{\sst{(7)}}}
\def\8{{\sst{(8)}}}
\def\sst#1{{\scriptscriptstyle #1}}

\def\im{{{\rm i}}}

\begin{document}
\begin{flushright}
\hfill{ \
MI-TH-1542\ \ \ \ }
\end{flushright}

\vspace{25pt}
\begin{center}
{\Large {\bf Evidence for the Holographic dual of ${\cal N}=3$ Solution in Massive Type IIA}
}

\vspace{30pt}

{\Large
Yi Pang\hoch{1} and Junchen Rong\hoch{2}
}

\vspace{10pt}

\hoch{1} {\it Max-Planck-Insitut f\"{u}r Gravitationsphysik (Albert-Einstein-Institut)\\
Am M\"{u}hlenberg 1, DE-14476 Potsdam, Germany}

\hoch{2} {\it George P. \& Cynthia Woods Mitchell  Institute
for Fundamental Physics and Astronomy,\\
Texas A\&M University, College Station, TX 77843, USA}

\vspace{10pt}

\vspace{20pt}

\underline{ABSTRACT}
\end{center}
\vspace{15pt}
We calculate the Kaluza-Klein spectrum of spin-2 fluctuations around the $\cN=3$ warped ${\rm AdS}_4\times M_6$ solution in massive IIA supergravity. This solution was conjectured to be dual to the $D=3$ $\cN=3$ superconformal ${\rm SU}(N)$ Chern-Simons matter theory with level $k$ and 2 adjoint chiral multiplets. The ${\rm SO}(3)_R\times{\rm SO}(3)_D$ isometry of the ${\cal N}=3$ solution is identified with the ${\rm SU}(2)_F\times {\rm SU}(2)_{\cal R}$ global symmetry of the dual ${\cal N}=3$ SCFT. We show that the ${\rm SO}(3)_R\times{\rm SO}(3)_D$ quantum numbers and the ${\rm AdS}$ energies carried by the BPS spin-2 modes match precisely with those of the spin-2 gauge invariant operators in the short multiplets of operators in the ${\cal N}=3$ SCFT. We also compute the Euclidean action of the ${\cal N}=3$ solution and the free energy of the ${\cal N}=3$ SCFT on $S^3$, in the limit $N\gg k$. Remarkably, the results show a complete agreement.

\thispagestyle{empty}

\pagebreak
\voffset=-40pt
\setcounter{page}{1}

\tableofcontents


\section{Introduction}
Since the seminal work of ABJM \cite{ABJM}, many progresses have been made in understanding the $D=3$ superconformal Chern-Simons matter theories and construction of their holographic duals. ABJM theory is a ${\rm U}(N)\times {\rm U}(N)$ Chern-Simons matter theory with explicit $\cN=6$ supersymmetry. In a certain regime, its gravity dual is the ${\rm AdS}_4\times\mathbb{CP}^3$ solution in IIA found long ago by \cite{Nilsson:1984bj}. One can consider deformations of ABJM theory by adding relevant superpotential or fundamental matter and its generalizations to quiver type gauge theories. These lead to new proposals of ${\rm AdS}_4/{\rm CFT}_3$ \cite{ABJ,Benna:2008zy,Jafferis:2008qz,Klebanov:2008vq,Ooguri:2008dk,Gaiotto:2009mv,Hohenegger:2009as,Gaiotto:2009tk,
Hikida:2009tp,Franco:2009sp,Klebanov0904,Gaiotto:2009yz,Petrini:2009ur,Ahn:2009et,Ahn:2009bq,Ammon:2009wc,Benini:2009qs,Jafferis:2009th}, in which the dynamics of the IR CFTs are still governed by certain superconformal Chern-Simons matter theories with more than one gauge groups. \footnote{Earlier proposals for superconformal duals to ${\rm AdS}_4$ have been made in \cite{Aharony:1998rm,Fabbri:1999hw,Ceresole9912,Billo0005,Billo00052}.}

In three dimensions, there exist also superconformal Chern-Simons matter systems with a single gauge group \cite{Schwarz:2004yj, Gaiotto:2007qi}. It is then natural to quest whether some of them possess supergravity duals. It was shown in \cite{Kao:1992ig,Kao:1995gf} that in general, $D=3$ Chern-Simons matter theories with a single gauge group admit at most $\cN=3$ supersymmetry. The spectra of BPS operators in $\cN=2,3$ superconformal ${\rm SU}(N)$ Chern-Simons matter theories with adjoint chiral matter fields were studied in \cite{Minwalla:2011ma}, which demonstrated that most of the superconformal Chern-Simons matter theories with a single gauge group did not have supergravity duals when the 't Hooft coupling is large, due to the presence of light protected higher spin operators and a Hagedorn growth in the specta \footnote{The analysis of \cite{Minwalla:2011ma} requires nontrivial $R$-symmetry, thus it does not apply to $\cN=1$ theories.}. Recently, a new $\cN=2$ warped ${\rm AdS_4}\times S^6$ solution in massive IIA was found in \cite{Guarino:2015jca} by uplifting the $\cN=2$ ${\rm U}(1)\times {\rm SU}(3)$ invariant stationary point in $D=4$ dyonic ISO(7) gauged maximal supergravity, which was first constructed in \cite{DallAgata:2011aa} and later proved to be a new model by \cite{ Dall'Agata:2014ita} (see also \cite{Guarino:2015qaa} for more details). The detailed derivation of the uplift formulas is given in \cite{Guarino:2015vca}. It was further proposed \cite{Guarino:2015jca} that this solution is holographically dual to a $D=3$ $\cN=2$ superconformal Chern-Simons matter theory with a single ${\rm SU}(N)$ gauge group, 3 adjoint chirals and a cubic superpotential. This theory can be viewed as the IR fixed point of a RG flow in the world volume theory of $N$ D2 branes with a Chern-Simons term induced by the Romans mass. The Chern-Simons level $k$ is related to the Romans mass $m$ by $m=k/(2\pi\ell_s)$. Some evidence for this proposal was provided by comparing the Euclidean action of the supergravity solution with the free energy of the dual CFT on $S^3$ \cite{Guarino:2015jca}. A generalization of this proposal has been put forward in \cite{Fluder:2015eoa}, by considering a system of $N$ D2 branes probing a generic Calabi-Yau threefold  singularity in massive IIA. In this case, the dual CFT was conjectured to be the IR fixed point of the low energy effective $\cN=2$ Chern-Simons quiver gauge theory. Some previous work on supersymmetric ${\rm AdS}_4\times M_6$ type solutions in (massive) type IIA supergravity can be found in \cite{Behrndt:2004km,Behrndt:2004mj,Nunez:2001pt,Lust:2004ig,Tomasiello:2007eq,Lust:2009mb,Rota:2015aoa,Varela:2015uca}.

The $D=4$ dyonic ISO(7) gauged maximal supergravity admits an $\cN=3$ stationary point \cite{Gallerati:2014xra}, which was uplifted to a warped ${\rm AdS}_4\times M_6$ solution in massive IIA \cite{paro}, utilizing formulas given in \cite{Guarino:2015jca}. The bosonic isometry of the $\cN=3$ solution is the ${\rm SO(3)}_R\times{\rm SO(3)}_D$ subgroup of SO(7) \cite{paro}, which can be characterized by starting from ${\rm SO}(4)\times {\rm SO}(3)_V \in {\rm SO}(7)$.  The factor ${\rm SO}(3)_R$ then comes from ${\rm SO}(4)\simeq{\rm SO}(3)_R\times {\rm SO}(3)_L$, and the factor ${\rm SO}(3)_D$ is the diagonal in ${\rm SO}(3)_L \times{\rm SO}(3)_V$\footnote{The ${\rm SO(3)}_R\times{\rm SO(3)}_D$ invariant subsector in the $\omega$-deformed family of SO(8)-gauged $\cN=8$ four-dimensional supergravities \cite{Dall'Agata:2012bb} was studied in \cite{paporo}.}. The $\cN=3$ solution in massive IIA was conjectured \cite{Guarino:2015jca} to be the gravity dual of the $\cN=3$ superconformal ${\rm SU}(N)$ gauged Chern-Simons matter theory with 2 adjoint chirals and a quartic superpotential studied in \cite{Gaiotto:2007qi}.
Interestingly, it was noticed in \cite{Minwalla:2011ma} that the BPS spectrum of the $\cN=3$ SCFT includes only states of spins $\leq$ 2, indicating the existence of a supergravity dual. The radius of curvature of the string frame metric in string units scales like $R/\ell_s\sim(N/k)^{\frac16}$, whilst the string coupling scales as $g_s\sim1/(N^{\frac16}k^{\frac56})$. Thus the supergravity description is valid when $N$ is much larger than $k$. Also, one can see that $g_s<\ell_s/R$, which is in agreement with previous observations that ``massive IIA cannot be strongly coupled" \cite{Aharony:2010af}, in other words, the string coupling must be small if the curvature is small.

In \cite{paro}, we made the first attempt of testing the conjectured duality between the $\cN=3$ solution in massive IIA and the $\cN=3$ superconformal ${\rm SU}(N)$ Chern-Simons matter theory. It was found that at the lowest level, the ${\rm SO(3)}_R\times{\rm SO(3)}_D$ quantum numbers and the AdS energies of the fluctuations around the $\cN=3$ background match with those of the short multiplets of gauge invariant operators in the $\cN=3$ SCFT. In this paper, we provide further evidence for this conjectured duality. We first perform an explicit analysis of the Kaluza-Klein (KK) spectrum of the spin-2 fields in ${\rm AdS}_4$. Solving the spin-2 spectrum in a warped AdS background with inhomogeneous internal space has been encountered in previous studies \cite{Klebanov0904,Ahn:2009bq,Ahn:2009et}\footnote{Earlier work on KK spectrum in a warped AdS background with homogeneous internal space can be found in \cite{Duff:1986hr,Gubser:1998vd,Fabbri:1999mk,Termonia:1999cs,Fre':1999xp,Merlatti:2000ed,Ceresole9910}.}, where it was found that the analytic solutions for the spin-2 fluctuations involve hypergeometric functions. Similarly, solutions for the spin-2 fluctuations around the $\cN=3$ solution in massive IIA also include hypergeometric functions. The squared masses of the gravitons obtained by imposing regularity of the solution, depend quadratically on the quantum numbers associated with ${\rm SO(3)}_R\times{\rm SO(3)}_D$ and ${\rm SO}(3)_V$. These results holds for both the BPS and non-BPS spin-2 excitations, which respectively belong to the short and long graviton multiplets of ${\rm OSP}(3|4)$. The ${\rm SO(3)}_R\times{\rm SO(3)}_D$ quantum numbers and the AdS energy carried by the BPS gravitons agree precisely with those of the gauge invariant spin-2 operators present in the spectrum of BPS operators of the dual $\cN=3$ SCFT. We then compare the free energy of the supergravity solution with that of the $\cN=3$ SCFT on $S^3$, in the limit $N\gg k$. Remarkably, the results show a complete agreement.

This paper is organized as follows. In Sec. \ref{spin2spectrum}, we give an explicit derivation of the KK spectrum for the spin-2 fluctuations around the $\cN=3$ solution in massive IIA. In Sec. \ref{dualspin2}, we proceed with identifying the CFT operators dual to the short graviton multiplets in the spectrum of the bulk theory. In Sec. \ref{freen}, we compute the Euclidean action of the supergravity background and the free energy of the $\cN=3$ superconformal ${\rm SU}(N)$ Chern-Simons matter theory on $S^3$. We conclude and discuss possible future directions in Sec. \ref{cd}.
\section{Kaluza-Klein spectrum of Spin-2 fluctuations}\label{spin2spectrum}
In solving the KK spectrum of the spin-2 modes, only the metric of the $\cN=3$ solution is needed.
The complete solution involving various form fields can be found in \cite{paro} (Its form in vielbein basis is given in the appendix of this paper.).
In terms of the seven auxiliary coordinates on $S^6$
\bea\label{munu}
\mu^1&=&\sin \xi \cos \theta _1  \cos \chi _1,\quad \mu^2=\sin \xi \cos \theta _1  \sin \chi _1,\quad
\mu^3=\sin \xi \sin \theta _1  \cos \psi ,\nn\\
\mu^4&=& \sin \xi \sin \theta _1\sin \psi ,\quad \nu^1=\cos \xi  \cos \theta _2 ,\quad  \nu^2=\cos \xi  \sin \theta _2  \cos \chi _2,\nn\\
\nu^3&=&\cos \xi  \sin \theta _2  \sin \chi _2,
\eea
which satisfy $\sum_{A=1}^{4}\mu^A\mu^A+\sum_{i=1}^{3}\nu^{i}\nu^{i}=1$, the metric of the $\cN=3$ solution in massive IIA takes the form \cite{paro}\footnote{In this section, we will choose $g=1$, $m=2$ for convenience. }
\be
d\hat{s}^2_{10}= \Delta^{-1}(\frac{3\sqrt{3}}{16}ds^2_{{\rm AdS}_4})+g_{mn} dy^m dy^n,
\ee
in which
\be
\Delta=3^{\ft98}2^{-\ft34}(\cos 2 \xi +3)^{-\ft18} \Xi^{-\ft14},\qquad \Xi=(24 \cos 2 \xi +3 \cos 4 \xi +37),
\ee
and the internal metric on the deformed $S^6$ is given as
\bea
g_{mn}dy^m dy^n&=\frac{3\sqrt{3}}{4}(\Delta{\Xi})^{-1} \bigg[&-\sin ^2 2 \xi  d\xi^2+8 (\cos 2 \xi +3) d\mu \cdot d\mu+4 (\cos 2 \xi +3)d\nu\cdot d\nu\nn\\&&+16 \mu^A \eta^i_{AB} d\mu^B\epsilon^{ijk}\nu^jd\nu^k-\frac{16}{\cos 2 \xi +3} (d\mu^A \eta^i_{AB} \mu^B \nu^i)^2\bigg],
\label{background}
\eea
where $\eta^i$'s are the generators of ${\rm SO}(3)_L$ embedded in ${\rm SO}(4)\simeq {\rm SO}(3)_R\times {\rm SO}(3)_L$,
\be
\eta^1=\left(
\begin{array}{cccc}
 0 & 0 & 0 & -1 \\
 0 & 0 & -1 & 0 \\
 0 & 1 & 0 & 0 \\
 1 & 0 & 0 & 0 \\
\end{array}
\right),\quad
\eta^2=\left(
\begin{array}{cccc}
 0 & 0 & 1 & 0 \\
 0 & 0 & 0 & -1 \\
 -1 & 0 & 0 & 0 \\
 0 & 1 & 0 & 0 \\
\end{array}
\right),\quad
\eta^3=\left(
\begin{array}{cccc}
 0 & -1 & 0 & 0 \\
 1 & 0 & 0 & 0 \\
 0 & 0 & 0 & -1 \\
 0 & 0 & 1 & 0
\end{array}
\right).
\ee
As explained in \cite{paro}, this solution respects an ${\rm SO}(3)_R\times {\rm SO}(3)_D$ symmetry, which is embedded in ISO(7) via the chain
\be
{\rm ISO}(7)\supset {\rm SO}(7)\supset {\rm SO}(3)_R\times {\rm SO}(3)_L \times {\rm SO}(3)_V\supset {\rm SO}(3)_R\times \big[{\rm SO}(3)_L \times {\rm SO}(3)_V\big]_D,
\ee
where $\big[{\rm SO}(3)_L \times {\rm SO}(3)_V\big]_D$ means the diagonal subgroup of ${\rm SO}(3)_L \times {\rm SO}(3)_V$, which we denote by ${\rm SO}(3)_D$.
The preserved $\cN=3$ supersymmetry transforms as ${\bf 3}$ of ${\rm SO}(3)_D$. Thus, via holography ${\rm SO}(3)_D$ should be identified with the $R$-symmetry of the dual SCFT, whilst ${\rm SO}(3)_R$ plays a role of the flavor symmetry. Later, we will solve the spin-2 fluctuations around the $\cN=3$ solution, and compare the results with the gauge invariant spin-2 BPS operators in the dual CFT. To avoid the confusion from the notation, we relabel the two ${\rm SO}(3)$ groups as
\be
{\rm SO}(3)_R\rightarrow {\rm SO}(3)_F,\qquad {\rm SO}(3)_D\rightarrow {\rm SO}(3)_{\cal R},
\ee
where now ${\rm SO}(3)_{\cal R}$ and ${\rm SO}(3)_F$ correspond to the $R$-symmetry group and the flavor symmetry group of the dual SCFT respectively.

We consider fluctuations of the metric around the $\cN=3$ background
\be
\hat{g}_{MN}\rightarrow \bar{g}_{MN}+\hat{h}_{MN}.
\ee
Similar to the cases studied in \cite{Klebanov0904,Ahn:2009et,Ahn:2009bq,Gubser:1997yh,Constable:1999gb}, applying the separation of variables to the transverse and traceless (with respect to the ${\rm AdS}_4$ metric $g_{4\mu\nu}$ without the warp factor) part of $\hat{h}_{\mu\nu}$,
\be
\hat{h}_{\mu\nu}=h_{\mu\nu}(x)Y(y),\,\quad {\nabla}_4^{\mu}h_{\mu\nu}=0,\,\quad {g}_4^{\mu\nu}h_{\mu\nu}=0,
\ee
we find that the spin-2 modes solving the homogenous linearized Einstein equation satisfy
\be
Y(y)L^{-2}_0(\Box_4+2)h_{\mu\nu}+h_{\mu\nu}{\cal O}Y(y)=0,\quad L^2_0=\frac{3\sqrt{3} }{16},
\label{spin2eq}
\ee
where $\Box_4$ is the Laplacian on the unit ${\rm AdS}_4$, and the operator ${\cal O}$ is given by
\bea
{\cal O}Y(y)&=&\frac{\Delta^{-1}}{\sqrt{-\bar{g}_{10}}}\partial_M(\sqrt{-\bar{g}_{10}}\bar{g}^{MN}\partial_N)Y(y)\nn\\
&=&\frac{1}{\sqrt{\mathring{g}_6}}\partial_m(\Delta^{-1}\sqrt{\mathring{g}_6}\bar{g}^{mn}\partial_n)Y(y),
\eea
where $\mathring{g}_6$ is the metric on the round $S^6$. The operator ${\cal O}$ can be written explicitly as
\be
L_0^2{\cal O}\equiv\widetilde{{\cal O}}=\frac{1}{2}\partial_{\xi}^2+\frac12(3 \cot\xi-2\tan\xi)\partial_{\xi}+\frac12\sec ^2\xi C_V+(2 \csc ^2\xi -1)C_F+\frac{C_{\cal R}-C_L}{2},
\label{spin0eq}
\ee
where $C_V$, $C_F$, $C_{\cal R}$ and $C_L$ are the quadratic Casimirs associated with ${\rm SO}(3)_V$, ${\rm SO}(3)_F$, ${\rm SO}(3)_{\cal R}$ and ${\rm SO}(3)_L$. When acting on scalars, these Casimirs can be expressed as bilinears of Lie derivatives associated with Killing vectors generating the corresponding SO(3). Killing vectors associated with ${\rm SO}(3)_V$ are given by
\be
\xi^{i}_V=\epsilon^{ijk}\nu^j\frac{\partial}{\partial \nu^k},
\ee
whilst Killing vectors associated with ${\rm SO}(3)_F$ and ${\rm SO}(3)_L$ take the form
\be
\xi^i_F=-\mu^A(T^i_F)_{AB}\frac{\partial}{\partial \mu^B}\,,\quad \xi^i_L=-\mu^A(T_L^i)_{AB}\frac{\partial}{\partial \mu^B}.
\ee
In the expressions above ,
\bea
T_F^1&=&-\frac{1}{2}(R^{12}-R^{34}),\quad
T_F^2=-\frac{1}{2}(R^{13}-R^{42}),\quad T_F^3=-\frac{1}{2}(R^{14}-R^{23}),\\
T^1_L&=&\frac{1}{2}(R^{12}+R^{34}),\quad T^2_L=\frac{1}{2}(R^{13}+R^{42}),
\quad T^3_L=\frac{1}{2}(R^{14}+R^{23}),
\eea
where the $R^{ij}$ are the ${\rm SO}(4)$ generators,
with $(R^{ij})_{ij}=-(R^{ij})_{ji}=1$, and all other elements equal to zero. Then the quadratic Casimirs are given by
\be
C_F={\cal L}_{\xi_F^{i}}{\cal L}_{\xi_F^{i}},\quad C_L={\cal L}_{\xi_L^{i}}{\cal L}_{\xi_L^{i}},\quad C_V={\cal L}_{\xi_V^{i}}{\cal L}_{\xi_V^{i}},\quad  C_{\cal R}=({\cal L}_{\xi_L^{i}}+{\cal L}_{\xi_V^{i}})({\cal L}_{\xi_L^{i}}+{\cal L}_{\xi_V^{i}}).
\ee
The harmonic function $Y(y)$ satisfies $\widetilde{\mathcal{O}}Y(y)=-m^2 Y(y)$ leading to
\be
(\Box_4+2)h_{\mu\nu}-m^2h_{\mu\nu}=0.
\ee
From the equation above, one can solve for the AdS energies carried by the spin-2 modes. For each $m^2$, we have
\be
E_0=\frac12(3+\sqrt{9+4m^2}).
\ee
To find eigenmodes for the operator $\widetilde{{\cal O}}$, it is useful to know the eigenfunctions of various Casimirs. We recall that
spin-0 harmonics on a round 6-sphere are characterized by $(n,0,0),\, n=1,2\cdots$ representations of ${\rm SO}(7)$ and also
 form a complete basis for smooth scalar functions
on manifold with $S^6$ topology. Thus, the decomposition of
the ${\rm SO}(7)$ harmonics under the ${\rm SO}(3)_F\times {\rm SO}(3)_{\cal R}$ subgroup
should give rise to a complete functional basis
on the internal space of the $\cN=3$ solution \eqref{background} which is a smooth deformation of $S^6$.
Since the ${\rm SO}(3)_F\times {\rm SO}(3)_{\cal R}$ subgroup is embedded in ${\rm SO}(7)$ via the chain
\be
{\rm SO}(7)\supset {\rm SO}(4)\times {\rm SO}(3)_V\simeq {\rm SO}(3)_F\times {\rm SO}(3)_L \times {\rm SO}(3)_V\supset{\rm SO}(3)_F\times {\rm SO}(3)_{\cal R},
\ee
we first branch the $(n,0,0)$ irrep under the ${\rm SO}(4)\times {\rm SO}(3)_V$ subgroup. This yields a sequence of
irreps of ${\rm SO}(4)\times {\rm SO}(3)_V$ of the form $(\ell,0)_{j_V}$, where $(\ell,0)$ correspond to the highest weights of the ${\rm SO}(4)$ irrep. Here $\ell,\, j_V$ are non-negative integers. Under the isomorphism ${\rm SO}(4)\simeq{\rm SO}(3)_F\times {\rm SO}(3)_L$, the highest weights $(\ell_1,\ell_2)$ of ${\rm SO}(4)$ are related to the isospins $(j_F,j_L)$ of ${\rm SO}(3)_F\times {\rm SO}(3)_L $ by
\be
j_F=\ft12(\ell_1+2\ell_2),\qquad j_L=\ft12\ell_1.
\ee
This means further branching of $(\ell,0)$ under ${\rm SO}(3)_F\times {\rm SO}(3)_L $ leads to a sequence of irreps with $j_F=j_L$.
The analysis above suggests that the eigenfunctions of the Casimirs $C_F,\,C_L,\,C_V$ should take the form
\be\label{eigenfun}
f(\xi)(\alpha_{A_1 A_2\cdots A_p }\prod^{p=2j_F}_{k=1} \widetilde{\mu}^{A_k})(\beta_{i_1 i_2\cdots i_q}\prod^{q=j_V}_{m=1} \widetilde{\nu}^{i_{m}}),\quad j_F\in\frac12\mathbb{Z}^+\cup\{0\},\quad j_V\in \mathbb{Z}^+\cup\{0\},
\ee
where
\be
\widetilde{\mu}^A= \frac{\mu^A}{\sin \xi},\quad A=1\cdots 4,\qquad
\widetilde{\nu}^i= \frac{\nu^i}{\cos \xi},\quad i=1\cdots 3,
\ee
and
$f(\xi)$ is a function of $\xi$ which cannot be determined by group theoretical analysis. Coefficients $\alpha_{A_1 A_2\dots A_p}$ and $\beta_{i_1 i_2\dots i_q}$ are totally symmetric, traceless with respect to their indices and transform according to the $(j_F,j_F,j_V)$ irrep of ${\rm SO}(3)_F\times {\rm SO}(3)_L \times {\rm SO}(3)_V $. Since ${\rm SO}(3)_{\cal R}$ is the diagonal of ${\rm SO}(3)_L \times {\rm SO}(3)_V$, the eigenfunctions of its Casimir can be obtained by decomposing the product of $\alpha_{A_1 A_2\cdots A_p}$ and $\beta_{i_1 i_2\cdots i_q}$ in terms of irreps of ${\rm SO}(3)_{\cal R}$ using Clebsch-Gordan coefficients. In the end, we achieve the mutual eigenfunctions for $C_F,\,C_L,\,C_V,\,C_{\cal R}$ labeled by the quantum numbers
\be
(j_F,j_F,j_V,j_{\cal R}),\quad j_{\cal R}=|j_V-j_F|,\cdots,j_V+j_F, \quad j_F\in\frac{1}{2}\mathbb{Z}^+\cup\{0\},\quad j_V\in \mathbb{Z}^+\cup\{0\}.
\label{SO(3)4}
\ee
For simplicity, we denote the eigenfunction obtained through the above procedure by the abstract symbol
\be
|\psi\rangle=|j_F,j_F,j_V,j_{\cal R}\rangle.
\ee
It satisfies
\bea
& C_F|\psi\rangle=c_F|\psi\rangle, &\qquad c_F=-j_F(j_F+1),\nn\\
& C_L|\psi\rangle=c_L|\psi\rangle, & \qquad c_L=-j_F(j_F+1),\nn\\
& C_V|\psi\rangle=c_V|\psi\rangle, &\qquad c_V=-j_V(j_V+1),\nn\\
& C_{\cal R}|\psi\rangle=c_{\cal R}|\psi\rangle, &\qquad c_{\cal R}=-j_{\cal R}(j_{\cal R}+1),
\eea
which also illustrates the normalization of the Casimirs.
Substituting the ansatz
\be
Y(y)=f(\xi)|j_F,j_F,j_V,j_{\cal R}\rangle,
\ee
into \eqref{spin0eq} and making the change of variable
\be
u=\cos^2\xi,\quad\widetilde{f}(u)\equiv f(\xi),
\ee
we arrive at an equation for $\widetilde{f}(u)$
\bea
(1-u)^2 u^2 \widetilde{f}''(u)&+&\frac12(7 u^3-10u^2+3 u)\widetilde{f}'(u)\label{diffeqn}\\
&+&\bigg(\frac14u (3u+1)c_F+\frac14(1-u)c_V+\frac14u(1-u)(c_{\cal R}+2m^2) \bigg)\widetilde{f}(u)=0.\nn
\eea
By a further change of variable
\be
\widetilde{f}(u)\equiv u^{j_V/2 }(1-u)^{j_F } H(u),
\ee
the equation above is brought to the form of a standard  hypergeometric differential equation
\be\label{Hypergeometric}
u(1-u)\frac{d^2H}{du^2}+\big(c-(a+b+1)u\big)\frac{d H}{du}-a b H(u)=0,
\ee
where the constants are given by
\bea
a&=&\frac{1}{4} (-\sqrt{12 j_F^2+12 j_F-4 j_{\cal R}^2-4 j_{\cal R}+8 m^2+25}+4 j_F+2 j_V+5),\nn\\
b&=&\frac{1}{4}  (\sqrt{12 j_F^2+12 j_F-4 j_{\cal R}^2-4 j_{\cal R}+8 m^2+25}+4 j_F+2 j_V+5 ),\nn\\
c&=&j_V+\frac{3}{2},
\eea
There are two independent solutions to the hypergeometric differential equation above
\be
 {}_2F_1(a,b,c,u),\, \quad {\rm and}\quad  u^{1-c}{}_2F_1(1+a-c,1+b-c,2-c,u).
\ee
The second solution should be discarded, since the corresponding $f(u)$ is singular at $u=0$. The first solution converges for $|u|<1$. It can be proved that for $(1-u)^{j_F} 2F_1(a,b,c,u)$ to be regular at $u=1$, the coefficient $a$ must be a non-positive integer. Regularity of the solution thus dictates the mass squared $m^2$ to depend on the quantum numbers quadratically
\be\label{spin2mass}
m^2=\frac{1}{2} \bigg(2 n(4 j_F+2 j_V+5)+4 j_F j_V+j_F^2+7 j_F+4 n^2+j_{\cal R}^2+j_{\cal R}+j_V^2+5 j_V\bigg),
\ee
where $n\in \mathbb{Z}^+\cup\{0\}$. A typical spin-2 excitation with ${\rm AdS}$ energy being an integer is given by
$n=0$, $j_F=0$ and $j_V=j_{\cal R}$, which leads to
\be
m^2=j_{\cal R}(j_{\cal R}+3), \quad E_0=j_{\cal R}+3,\quad j_{\cal R}\in \mathbb{Z}^+\cup\{0\}.
\label{shortg}
\ee
It should be noted that gravitons with the same ${\rm SO}(3)_F\times{\rm SO}(3)_{\cal R}$ quantum numbers and ${\rm AdS}$ energies appear in the short graviton multiplet ${\rm DS}(2,j_{\cal R}+3/2,j_{\cal R}|3)$ of ${\rm OSP}(3|4)$ \cite{Fre':1999xp,Freedman1984}. Since the supergravity background preserves $\cN=3$ superconformal symmetry, the spin-2 states \eqref{shortg} must form complete ${\rm DS}(2,j_{\cal R}+3/2,j_{\cal R}|3)$ multiplets together with other lower spin states with proper quantum numbers and ${\rm AdS}$ energies. The spin-2 states \eqref{shortg} are singlets with respect to ${\rm SO}(3)_F$, which means all the states belonging to the short graviton multiplets are singlets of the flavor symmetry. On the CFT side, the spectrum of BPS operators in the $\cN = 3$ superconformal ${\rm SU}(N)$ Chern-Simons-matter theory with 2 adjoint chirals has been studied by \cite{Minwalla:2011ma}. It was shown that the short multiplets ${\rm DS}(2,j_{\cal R}+3/2,j_{\cal R}|3)$ composed by gauge invariant operators are singlets of the flavor symmetry. Therefore, our results demonstrate a perfect matching between the short graviton multiplets in the KK spectrum of fluctuations around the $\cN=3$ vacuum in massive IIA and the short multiplets involving spin-2 operators in the $\cN=3$ superconformal ${\rm SU}(N)$ Chern-Simons matter theory with two adjoint chirals.

A list of the bulk spin-2 states labeled by their quantum numbers is given in Table \ref{N=3longgraviton}, from which one can see that the spectrum includes long graviton multiplets with rational dimensions. This feature has been observed for other M-theory and string theory backgrounds \cite{Gubser:1998vd,Ceresole9910,Ceresole9912,Billo0005,Billo00052,Klebanov0904}. A class of long multiplets with rational dimensions was termed as the ``shadow'' multiplets \cite{Billo00052}. From the bulk point of view, shadowing mechanism is related to the fact that the same harmonics also appear in other fields belonging to short multiplets. In the spectrum obtained here, the long graviton labeled by $(j_F,j_V,j_{\cal R},n)=(1,r,r,0)$ carries $E_0=r+4$. The corresponding long graviton multiplets are shadows of vector multiplets.
\begin{table*}\centering
\ra{1.1}
{\small
\begin{tabular}{@{}cccccccccc@{}}
\toprule
\hline$(j_F,j_V,j_{\cal R})$&{$n$} && $m^2$ && $E_0$ &S(hort)/L(ong)
 \\ \midrule
$(0,~0,~0)$
 &0 && 0 && 3 & S\\ \midrule
$(0,~0,~0)$  &1&& 7  && $\frac12(3+\sqrt{37})$& L\\ \midrule
$(0,~0,~0)$  &2&& 18  && 6 & L\\ \midrule
$(0,~1,~1)$
  &0&& 4  && 4 & S\\\midrule
$(0,~1,~1)$  &1&& 13  && $\frac12(3+\sqrt{61})$ & L\\\midrule
$(0,~1,~1)$ &2&& 26  && $\frac12(3+\sqrt{113})$ & L\\\midrule
$(\frac12,~0,~\frac12)$
  &0&& $\frac94$  && $\frac12(3+3\sqrt{2})$ & L\\ \midrule
 $(\frac12,~0,~\frac12)$ &1&& $\frac{45}{4}$ &&$\frac12(3+3\sqrt{6})$  & L\\ \midrule
 $(\frac12,~0,~\frac12)$ &2&& $\frac{97}4$  && $\frac12(3+\sqrt{106})$ & L\\\midrule
$(\frac12,~1,~\frac12)$
  &0&& $\frac{25}4$  && $\frac12(3+\sqrt{34})$ & L\\ \midrule
 $(\frac12,~1,~\frac12)$ &1&& $\frac{69}{4}$ &&$\frac12(3+\sqrt{78})$  & L\\ \midrule
 $(\frac12,~1,~\frac12)$ &2&& $\frac{129}4$  && $\frac12(3+\sqrt{138})$ & L\\\midrule
$(\frac12,~1,~\frac32)$
  &0&& $\frac{31}4$  && $\frac12(3+2\sqrt{10})$ & L\\ \midrule
 $(\frac12,~1,~\frac32)$ &1&& $\frac{75}{4}$ &&$\frac12(3+2\sqrt{21})$  & L\\ \midrule
 $(\frac12,~1,~\frac32)$ &2&& $\frac{135}4$  && $\frac{15}{2}$ & L\\\midrule
$(1,~0,~1)$
  &0&& 5  && $\frac12(3+\sqrt{29})$ &L\\\midrule
$(1,~0,~1)$  &1&& 16  && $\frac12(3+\sqrt{73})$ & L\\\midrule
$(1,~0,~1)$  &2&& 31  && $\frac12(3+\sqrt{133})$ & L\\\midrule
$(1,~1,~0)$
  &0&& 9  && $\frac12(3+3\sqrt{5})$ & L\\\midrule
$(1,~1,~0)$  &1&& 22  && $\frac12(3+\sqrt{97})$ & L\\\midrule
$(1,~1,~0)$  &2&& 39  && $\frac12(3+\sqrt{165})$ & L\\\midrule
$(1,~1,~1)$
  &0&& 10  && $5$ & L\\\midrule
$(1,~1,~1)$  &1&& 23  && $\frac12(3+\sqrt{101})$ & L\\\midrule
$(1,~1,~1)$  &2&& 40  && 8 & L\\\midrule
 \bottomrule
\end{tabular}
}\normalsize
\caption{An incomplete list of the KK spectrum of spin-2 states. The ``Short'' and ``Long'' refer to the short and long multiplets which the spin-2 states belong to. Here we remind the reader that $j_{\cal R}=|j_V-j_F|,\cdots,j_V+j_F$, $j_F\in\frac{1}{2}\mathbb{Z}^+\cup\{0\},\quad j_V\in \mathbb{Z}^+\cup\{0\}$.}
\label{N=3longgraviton}
\end{table*}
\section{Matching short spin-2 multiplets}\label{dualspin2}
We can go further to identify the CFT operators dual to the bulk spin-2 modes satisfying \eqref{shortg}. Before doing so, we first briefly review the $\cN=3$ superconformal ${\rm SU}(N)$ Chern-Simons matter theory with 2 adjoint chirals. In $D=3$, the $\cN=3$ superconformal symmetry must have an ${\rm SO}(3)$ $R$-symmetry. As a consequence, in the vector multiplet, the fermions are a triplet and a singlet of ${\rm SU}(2)_{\cal R}$, and the three scalar fields are a triplet (as are the three auxiliary fields). In the chiral multiplets, all fields are doublets of ${\rm SU}(2)_{\cal R}$. Let $a,b,\cdots$ be the ${\rm SU}(2)_{\cal R}$ indices and $I,J,\cdots$ be the indices for the fundamental representation of ${\rm SU}(2)_F$. The components of the matter fields are scalars $q^{Ia}$ and fermions $\psi^{Ia}$ subject to the reality conditions\footnote{We will generally adhere to the notations and conventions of \cite{Gaiotto:2007qi}}
\be
(q)^{\dagger}_{Ia}=\epsilon_{IJ}\epsilon_{ab}q^{Jb},\qquad \bar{\psi}_{Ia}=\im \sigma_2\epsilon_{IJ}\epsilon_{ab}\psi^{Jb}.
\ee
The Lagrangian of $\cN=3$ superconformal ${\rm SU}(N)$ Chern-Simons matter theory with manifest ${\rm SU}(2)_{\cal R}\times{\rm SU}(2)_F$ symmetry can be written as \cite{Gaiotto:2007qi}
\bea
{\cal L}&=&\frac{k}{4\pi}\big[CS(A)+{\rm Tr}(D^{ab}s_{ab}-\frac12\chi^{ab}\chi_{ab}+\chi\chi+\frac16s^{ab}[s_{bc},s^{c}_{~a}])\big]\nn\\
&&+\frac12|\nabla_{\mu}q_{Ia}|^2+\frac12q_{Ia}D^{ab}q^{I}_{~b}-\frac14|s_{ab}q^{Ic}|^2\nn\\
&&+\frac{\im}2\psi_{Ia}\gamma^{\mu}\nabla_{\mu}\psi^{Ia}-\frac12\psi_{I}^{~a}s_{ab}\psi^{Ib}+\im q_I^{~a}\chi_{ab}\psi^{Ib}+\im q_{Ia}\chi\psi^{Ia}.
\label{N=3lag}
\eea
We will formulate the short spin-2 multiplets in the $\cN=3$ SCFT in terms of $\cN=2$ superfields, due to the complicity of the $\cN=3$ superspace.
The decomposition of ${\rm Osp}(3|4)$ multiplets under ${\rm Osp}(2|4)$ was studied in \cite{Fre':1999xp}. For instance, the decomposition of a generic $\cN=3$ long spin-2 multiplet is given by\footnote{Slightly different from the notation used in \cite{Fre':1999xp}, here we use $\Delta_0$ to denote the lowest conformal weight in an ${\rm Osp}(3|4)$ multiplet instead of $E_0$, since the latter has been used for the AdS energy of the graviton.}
\bea
{\rm DS}(2,\Delta_0>J_0+3/2,J_0|3)&&\rightarrow \bigoplus_{y=-J_0}^{J_0} {\rm DS}(2,\Delta_0+1/2,y|2) \oplus \bigoplus_{y=-J_0}^{J_0} {\rm DS}(3/2,\Delta_0,y|2)\\
&&\oplus \bigoplus_{y=-J_0}^{J_0} {\rm DS}(3/2,\Delta_0+1,y|2)\oplus \bigoplus_{y=-J_0}^{J_0} {\rm DS}(1,\Delta_0+1/2,y|2).\nn
\label{N=3toN=2L}
\eea
Specific to an $\cN=3$ spin-2 multiplet, the multiplet is shortened when $\Delta_0=J_0+3/2$ \cite{Freedman1984}. Accordingly, the short multiplet decomposes into less $\cN=2$ multiplets
\be
{\rm DS}(2,J_0+3/2,J_0|3)\rightarrow \bigoplus_{y=-J_0}^{J_0} {\rm DS}(2,J_0+2,y|2) \oplus \bigoplus_{y=-J_0}^{J_0} {\rm DS}(3/2,J_0+3/2,y|2),
\label{N=3toN=2S}
\ee
where ${\rm DS}(2,J_0+2,J_0|2)$ and ${\rm DS}(3/2,J_0+3/2,J_0|2)$ denote the $\cN=2$ short graviton and short gravitino multiplets respectively.
 The detailed structure of ${\rm Osp}(2|4)$ supermultiplets can be found in \cite{Nicolai:1985hs,Klebanov:2008vq}. In $\cN=2$ notation,
the scalars $q^{Ia}$ can be parametrized by two complex scalar fields $(Z^1, Z^2)$
\be
q^{I1}=(Z^1,Z^2),\quad q^{I2}=(-\bar{Z}^2,\bar{Z}^1 ).
\ee
 The ${\rm U}(1)$ $R$-symmetry of $\cN=2$ supersymmetry is embedded in ${\rm SU}(2)_{\cal R}$ in such a way that the ${\rm U}(1)$ charges carried by $Z^1$ and $Z^2$ are both equal to $\frac12$ as required by the $\cN=3$ superconformal symmetry.
We introduce two chiral superfields ${\cal Z}^1$ and ${\cal Z}^2$ whose lowest components are given by scalars $Z^1$ and $Z^2$. Using the stress tensor superfield $\mathcal{T}^{(0)}_{\alpha\beta}$
\be
{\cal T}^{(0)}_{\alpha\beta}=\bar{{\cal D}}_{(\alpha}\bar{\mathcal{Z}}_i {\cal D}_{\beta)}\mathcal{Z}^i+\im \bar{\mathcal{Z}}_i \overleftrightarrow{\partial}_{\alpha\beta}\mathcal{Z}^i,
\ee
the short spin-2 multiplet ${\rm DS}(2,J_0+2,J_0|2)$ of gauge invariant operators can be expressed as
\be
{\rm Tr}\big[{\cal T}^{(0)}_{\alpha\beta}({\cal Z}^k{\cal Z}_k)^{J_0}\big],
\ee
which possesses the correct conformal dimension and ${\rm U}(1)_R$ charge and also satisfies the shortening condition. The spin-2 component in this multiplet has conformal dimension $E_0=3+J_0$. The $\cN=2$ short gravitino multiplet ${\rm DS}(2,J_0+3/2,J_0|2)$ can be realized as \cite{Klebanov:2008vq}
\be
{\rm Tr}\big[{\cal Z}^{k}({\cal D}_{\alpha} {\cal Z}_{k})({\cal Z}^i {\cal Z}_i)^{J_0}\big],
\ee
where the conformal dimension and the ${\rm U}(1)_R$ charge associated with the supercovariant derivative ${\cal D}_{\alpha}$ are $\frac12$ and $-1$.
Other $\cN=2$ long multiplets present in the decomposition \eqref{N=3toN=2S} can be obtained by employing a sequence of ${\rm SU}(2)_{\cal R}$ transformations on
the short graviton and gravitino multiplets. We summarize the results in Table \ref{N=3graviton}.

\begin{table}[h]
\begin{center}
\setlength{\arrayrulewidth}{.07em}

\begin{tabular}{|l|l|}
\hline
$\cN=2$ multiplet &  Operator\\
\hline ${\rm DS}(2,J_0+2,J_0|2)$&${\rm Tr}\big[{\cal T}^{(0)}_{\alpha\beta}(\mathcal{Z}^i\mathcal{Z}_i)^{J_0}\big]$\\\hline
${\rm DS}(2,J_0+2,J_0-1|2)$&$\sum\limits_{n=1}^{J_0}{\rm Tr}\big[{\cal T}^{(0)}_{\alpha\beta}({\cal Z}^{i_1}{\cal Z}_{i_1})\cdot\cdot(\bar{{\cal Z}}^{i_n}{\cal Z}_{i_n}+{\cal Z}^{i_n}\bar{{\cal Z}}_{i_n})\cdot\cdot({\cal Z}^{i_{J_0}}{\cal Z}_{i_{J_0}})\big]$
\\\hline
\vdots&\vdots\\\hline
 ${\rm DS}(2,J_0+2,-J_0|2)$& ${\rm Tr}\big[{\cal T}^{(0)}_{\alpha\beta}(\bar{\mathcal{Z}}^k\bar{\mathcal{Z}}_k)^{J_0}\big]$\\\hline
${\rm DS}(3/2,J_0+3/2,J_0|2)$& ${\rm Tr}\big[{\cal Z}^{k}(D_{\alpha} {\cal Z}_{k})({\cal Z}^i {\cal Z}_i)^{J_0}\big]$\\\hline
\multirow{2}{*}{${\rm DS}(3/2,J_0+3/2,J_0-1|2)$}&${\rm Tr}\big[\bar{{\cal Z}}^{i}(D_{\alpha} {\cal Z}_{i})({\cal Z}^j {\cal Z}_j)^{J_0}+{\cal Z}^{i}(\overline{D_{\alpha} {\cal Z}}_{i})({\cal Z}^j {\cal Z}_j)^{J_0}$\\&+$\sum\limits_{n=1}^{J_0}{\cal Z}^{k}(D_{\alpha} {\cal Z}_{k})({\cal Z}^{i_1}{\cal Z}_{i_1})\cdot\cdot(\bar{{\cal Z}}^{i_n}{\cal Z}_{i_n}+{\cal Z}^{i_n}\bar{{\cal Z}}_{i_n})\cdot\cdot({\cal Z}^{i_{J_0}}{\cal Z}_{i_{J_0}})\big]$\\\hline
\vdots&\vdots\\\hline
${\rm DS}(3/2,J_0+3/2,-J_0|2)$& ${\rm Tr}\big[\bar{{\cal Z}}^{k}(\bar{D}_{\alpha} \bar{{\cal Z}}_{k})(\bar{{\cal Z}}^i \bar{{\cal Z}}_i)^{J_0}\big]$\\\hline
\end{tabular}
\caption{In this table, we show how an $\cN=3$ short graviton multiplet ${\rm DS}(2,3/2+J_0,J_0|3)$ decomposes into $\cN=2$ graviton and gravitino multiples. It should be noted that all the $\cN=2$ multiplets are singlets of ${\rm SU}(2)_F$. Since all the superfields are in the adjoint representation of ${\rm SU}(N)$, the single trace operators are defined up to certain ordering. }
\label{N=3graviton}
\end{center}
\end{table}
On the bulk side, the harmonics associated with the spin-2 states in the short graviton multiplet ${\rm DS}(2,j_{\cal R}+3/2,j_{\cal R}|3)$ take the form
\be\label{shorteigen}
Y(y)\sim \beta_{i_1 i_2\cdots i_q}\prod^{q=j_{\cal R}}_{m=1} {\nu}^{i_{m}},
\ee
 where $\nu^i$ is defined in \eqref{munu}. By a comparison with the short spin-2 multiplets on the CFT side listed in Table \ref{N=3graviton}, we can make the identification
 \be
 (\nu^1+\im\nu^2,\nu^1-\im\nu^2,\nu^3)\rightarrow (\mathcal{Z}^k\mathcal{Z}_k,\bar{\mathcal{Z}}^k\bar{\mathcal{Z}}_k,\bar{\mathcal{Z}}^k\mathcal{Z}_k+\mathcal{Z}^k\bar{\mathcal{Z}}_k), \quad \bar{\mathcal{Z}}_k\equiv (\mathcal{Z}^k)^*,
 \label{id1}
 \ee
 where the ${\rm SU}(2)_F$ indices are raised and lowered by $\epsilon^{ij}$ and $\epsilon_{ij}$.
It can be checked that the right hand side of the above expression resides in the same representation of ${\rm SO}(3)_F\times{\rm SO}(3)_{\cal R}$ as $\nu^i$. This identification is somewhat counterintuitive in the sense that usually bulk coordinates transverse to the brane are related to the scalar fields on the brane linearly. In fact, the mapping \eqref{id1} can be made linear by using the fact that the equation of $D^{ab}$ derived from the Lagrangian \eqref{N=3lag} implies
\be
s_{ab}\sim[q_{I(a},q^I_{~b)}].
\ee
Therefore, $\nu^i$ can also be identified as $s_{ab}$. This identification is more natural for the reason below. The ${\cN} = 3$ SCFT studied here can be viewed as the IR limit of the ${\cN} = 4$ supersymmetric Yang-Mills gauge theory with the same matter content, and deformed by an $\cN=3$ Chern-Simons term \cite{ABJM}. Due to the presence of the Chern-Simons term, all the dynamical fields in the $\cN = 4$ vector multiplet become massive and may be integrated out at energy scale much lower than the mass scale, leading to the action \eqref{N=3lag}. In the supersymmetric Yang-Mills theory, scalars $s_{ab}$ are dynamical. Together with $Z^1$ and $Z^2$, the seven dynamical scalars, comprise the pullback of the seven coordinates in the directions transverse to $N$ coincident D2 branes. Based on the symmetry property with respect to ${\rm SO}(3)_F\times{\rm SO}(3)_{\cal R}$, one can also identify $\mu^A$ with $q^{Ia}$. However, to test this identification requires the knowledge of the harmonics associated with fluctuations of spins $<$ 2, which is beyond the scope of this paper and deserves future investigation.

\section{Matching free energies}\label{freen}
We now turn to compute the free energy of the $\cN=3$ solution in massive IIA and that of the $\cN=3$ CFT on $S^3$ in the limit $N\gg k$. This limit ensures the validity of the supergravity approximation of massive IIA string and also simplifies the expression for the CFT free energy obtained from a saddle point approximation. We show these two quantities agree with each other precisely. In this section, we recover the $g$ and $m$ dependence of the supergravity solution as they are important for the comparison.

The number of massive D2 branes which is equal to the rank of the gauge group is determined by the quantized Page charge \cite{Marolf:2000cb,Aharony:2009fc}.
\be
\int_{S^6}\tilde{F}_{(6)}=\int_{S^6}e^{\frac{1}{2}\phi}\hat{*}F_{(4)}+A_{(2)}\wedge dA_{(3)}+\frac{1}{6}m A_{(2)}\wedge A_{(2)}\wedge A_{(2)}=-(2\pi \ell_s)^5 N.
\label{fluxqua}
\ee
Plugging the $\cN=3$ solution \cite{paro},
we get
\be \label{flux1}
\frac{1}{(2\pi \ell_s)^5 g^5}\frac{16 \pi ^3}{3}=N.
\ee
On the other hand, the gravitational free energy is inversely proportional to the effective $D=4$ Newton's constant
\be
F_{\rm gravity}=\frac{\pi\ell^2}{2 G_4},\quad \ell^2=\frac{3\sqrt3}{16}g^{-7/3}(m/2)^{1/3},
\ee
where $\ell$ is the radius of ${\rm AdS}_4$ and the effective $D=4$ Newton's constant is related to the string length by
\be
\frac1{16\pi G_4}=\frac{2\pi}{(2\pi \ell_s)^8 g^6}{\rm Vol}(S^6).
\ee
In the equation above , ${\rm Vol}(S^6)=\frac{16}{15}\pi^3$ is the area of a unit $S^6$.
Finally, using the relation between the Romans mass parameter and the induced Chern-Simons level \cite{Gaiotto:2009mv}
\be
m=F_{(0)}=\frac{k}{2\pi \ell_s},
\ee
we can express the free energy of the $\cN=3$ supergravity solution in terms of $k$ and $N$
\be
F_{\rm gravity}=\frac{9\pi}{40} 3^{1/6} k^{1/3} N^{5/3}.
\ee
Various field strengths
\bea
&&F_{(2)}=dA_{(1)}+mA_{(2)}, \quad F_{(3)}=dA_{(2)},\nn\\
&&F_{(4)}=dA_{(3)}+A_{(1)}\wedge dA_{(2)}+\frac{ m}{2} A_{(2)}\wedge A_{(2)},
\eea
are invariant under the gauge transformations
\bea
&&\quad A_{(1)}\rightarrow A_{(1)}-d\Lambda_{(0)}-m\Lambda_{(1)},\quad A_{(2)}\rightarrow  A_{(2)}+d\Lambda_{(1)},\nn\\
&& A_{(3)}\rightarrow A_{(3)}+d\Lambda_{(3)}-d\Lambda_{(0)}\wedge A_{(2)}-m \Lambda_{(1)}\wedge A_{(2)}-\frac{m}2\Lambda_{(1)}\wedge d\Lambda_{(1)}.
\eea
However, apparently the Page charge \eqref{fluxqua} is not gauge invariant. The charge density in \eqref{fluxqua} is shifted by terms of the form
\be
\tilde{F}_{(6)}\rightarrow \tilde{F}_{(6)}+d\Lambda_{(0)}\wedge d(A_{(2)}\wedge A_{(2)})+d\Lambda_{(1)}\wedge dA_{(3)}-\frac{m}{2}\cdot d(\Lambda_{(1)}\wedge A_{(2)}\wedge A_{(2)}).
\ee
Because the internal space of the $\cN=3$ solution is a smooth deformation of $S^6$, its topology must be the same as that of $S^6$. We know $H_1(S^6)=0$ and $H_2(S^6)=0$, hence $\Lambda_{(0)}$ in $d\Lambda_{(0)}$ and $\Lambda_{(1)}$ in $d\Lambda_{(1)}$ are globally defined. We also checked that all the gauge potentials are globally defined too. Thus, the Page charge (\ref{fluxqua}) is in fact gauge invariant.

The free energy for a generic $D=3$ $\cN=2$ Chern-Simons quiver theory on $S^3$ has been investigated in \cite{Fluder:2015eoa}. For a theory with $G$ ${\rm SU}(N)$ gauge fields and some number of bifundamental and adjoint chiral multiplets, giving the $R$-charge spectrum $\{\Delta^I: I\in\, {\rm matter~fields}\}$, in the limit $N\gg k$, the free energy reads
\be
F=\frac{3\sqrt{3}\pi}{20\cdot 2^{1/3}}\bigg\{G+\sum_{I\in\, {\rm matter~fields}}(1-\Delta^I)\big(1-2 (1-\Delta^I)^2\big)\bigg\}^{2/3}k^{1/3}N^{5/3}.
\label{parfun}
\ee
In $\cN=2$ notation, the $\cN=3$ theory possesses a single ${\rm SU}(N)$ gauge group and two chiral multiplets. The $\cN=3$ superconformal symmetry determines the value of the ${\rm U}(1)_R$-charge carried by the lowest component in the chiral multiplet to be $\frac12$. Therefore, from \eqref{parfun}, we obtain
\be
F_{\rm SCFT}^{\cN=3}=\frac{9\pi}{40} 3^{1/6} k^{1/3} N^{5/3},
\ee
which agrees precisely with the free energy of the $\cN=3$ solution. Comparing the free energy of the $\cN=3$ SCFT with that of the $\cN=2$ SCFT \cite{Guarino:2015jca}, we see
\be
F_{\rm SCFT}^{\cN=3}<F_{\rm SCFT}^{\cN=2},
\ee
which is compatible with the expectation from F-theorem \cite{Jafferis:2011zi} in $D=2+1$ dimensions, since the ${\cal N}=3$ SCFT can be arrived via an RG-flow starting from the ${\cal N}=2$ SCFT in the UV \cite{Gaiotto:2007qi}.

\section{Discussions and conclusions}\label{cd}
In this paper we use direct KK reduction to calculate the spectrum of spin-2 modes around the $\cN=3$ warped ${\rm AdS}_4\times M_6$ solution in massive IIA supergravity. This solution was conjectured to be dual to the $\cN=3$ superconformal ${\rm SU}(N)$ Chern-Simons matter theory with 2 adjoint chirals \cite{Guarino:2015vca}. The ${\rm SO}(3)_R\times{\rm SO}(3)_D$ isometry of the $\cN=3$ solution is identified with the ${\rm SU}(2)_F\times {\rm SU}(2)_{\cal R}$ global symmetry of the dual $\cN=3$ SCFT. The KK spectrum of spin-2 modes includes both BPS states and non-BPS states. The former belong to the short graviton multiplets ${\rm DS}(2,j_{\cal R}+3/2,j_{\cal R}|3)$ of ${\rm OSP}(3|4)$ whilst the latter fall into the long graviton multiplets. The ${\rm SO}(3)_R\times{\rm SO}(3)_D$ quantum numbers and the ${\rm AdS}$ energies carried by the BPS spin-2 excitations match precisely with those carried by the spin-2 gauge invariant operators in the spectrum of BPS operators of the $\cN=3$ SCFT. The harmonics associated with the BPS spin-2 modes in the bulk also provide clues for the expressions of the spin-2 gauge invariant operators on the boundary CFT. It would be interesting to extend the analysis performed in the paper to KK excitations of different ${\rm AdS}_4$ spins. This is made harder by the relatively small amount of symmetry in this background and by fairly complicated expressions for the background metric and various form fluxes.

We performed a further check of the conjectured duality by comparing the free energy of the supergravity solution with that of the $\cN=3$ superconformal ${\rm SU}(N)$ Chern-Simons matter theory with 2 adjoint chirals. We show that these two quantities agree with each other. We expect this agreement still holds when the AdS vacuum is replaced by an AdS black hole and the dual CFT is at non-zero temperature. However, to confirm this expectation, one needs black hole solutions asymptotic to the $\cN=3$ warped AdS vacuum in massive IIA. Thus it should be interesting to look for such black hole solutions. Another interesting solution is the $\cN=2$ domain wall solution interpolating between the $\cN=2$ solution \cite{Guarino:2015jca} and the $\cN=3$ solution \cite{paro}, since the CFT dual to the $\cN=2$ solution flows to the one dual to the $\cN=3$ solution when one of the three chiral multiplets acquires a mass term \cite{Gaiotto:2007qi}.
\appendix
\section{${\cal N}=3$ Solution in vielbein basis }
\label{vielbein}
In this appendix, we present a vielbein system for the $\cN=3$ solution which inherits the ${\rm SO}(3)_R\times{\rm SO}(3)_D$ isometry of the metric. This vielbein system should be useful for solving the spectrum of fermionic fluctuations around the $\cN=3$ vacuum. We first introduce a few definitions
\bea
\ell^i_1&\equiv&\partial_{\phi_2}(\frac{\nu^i}{\cos\xi}),\quad \vec{\ell}_1=(0,\, -\sin\phi_2,\, \cos\phi_2),\nn\\
\vec{\ell}^i_2&\equiv&\partial_{\theta_2}(\frac{\nu^i}{\cos\xi}),\quad \vec{\ell}_2=(-\sin\theta_2,\, \cos\theta_2\cos\phi_2,\, \cos\theta_2\cos\phi_2),\nn\\
\vec{\ell}^i_3&\equiv&\frac{\nu^i}{\cos\xi},\qquad\quad \vec{\ell}_3=(\cos\theta_2,\, \sin\theta_2\cos\phi_2,\, \sin\theta_2\sin\phi_2),\nn\\
{\cal K}^{i}&\equiv&\mu^A \eta^i_{AB} d\mu^B,\quad {\cal J}^i\equiv\epsilon^{ijk}\nu^jd\nu^k.
\eea
From now on, we will set $g=1,\,m=2$ for brevity. The dependence on generic values of $g,\,m$ can be recovered using the scaling symmetry \cite{paro}. A choice of the ${\rm SO}(3)_R\times{\rm SO}(3)_D$ invariant vielbein system is the following
\bea
\hat{e}^0&=&\ft{3^{\ft34}}4\Delta^{-\ft12}e^0,\quad \hat{e}^1=\ft{3^{\ft34}}4\Delta^{-\ft12}e^1,\nn\\
\hat{e}^2&=&\ft{3^{\ft34}}4\Delta^{-\ft12}e^2,\quad \hat{e}^3=\ft{3^{\ft34}}4\Delta^{-\ft12}e^3,\nn\\
\hat{e}^4&=&\frac{3^{\ft34}}{\sin\xi}\sqrt{\frac{2(3+\cos 2\xi)}{\Xi\Delta}}\ell_1^i({\cal K}^i+\frac{\sin^2\xi}{3+\cos 2\xi}{\cal J}^i),\nn\\
\hat{e}^5&=&\frac{3^{\ft34}}{\sin\xi}\sqrt{\frac{2(3+\cos 2\xi)}{\Xi\Delta}}\ell_2^i({\cal K}^i+\frac{\sin^2\xi}{3+\cos 2\xi}{\cal J}^i),\nn\\
\hat{e}^6&=&\frac{3^{\ft34}}{\sin\xi\sqrt{2(3+\cos 2\xi)\Delta}}\ell_3^i{\cal K}^i,\nn\\ \hat{e}^7&=&\frac{3^{\ft34}}{2\cos\xi\sqrt{\Delta}}\ell^i_1(\frac{\nu^id\xi}{\sqrt2}+\frac{{\cal J}^i}{\sqrt{3+\cos 2\xi}}),\nn\\
\hat{e}^8&=&\frac{3^{\ft34}}{2\cos\xi\sqrt{\Delta}}\ell^i_2(\frac{\nu^id\xi}{\sqrt2}+\frac{{\cal J}^i}{\sqrt{3+\cos 2\xi}}),\nn\\ \hat{e}^9&=&\frac{3^{\ft34}}{2\cos\xi\sqrt{\Delta}}\ell^i_3(\frac{\nu^id\xi}{\sqrt2}+\frac{{\cal J}^i}{\sqrt{3+\cos 2\xi}}),
\eea
in which $e^0,\cdots,e^3$ comprise a vierbein system of the ``unit" $AdS_4$. In the vielbein basis given above, the field strengths take the form
\bea
\hat{F}_2&=&\bigg[\frac{16\Delta\cos\xi\sin^2\xi}{3\sqrt3(3+\cos2\xi)^2}\hat{e}^4\wedge\hat{e}^5+ \frac{4\Xi^{\ft12}\Delta\sin\xi(5+\cos2\xi)}{3\sqrt6(3+\cos2\xi)^2}(\hat{e}^4\wedge\hat{e}^8-\hat{e}^5\wedge\hat{e}^7)\nn\\
&&+\frac{32\Delta\sin^2\xi}{3\sqrt3(3+\cos2\xi)^{3/2}}\hat{e}^6\wedge\hat{e}^9
-\frac{16\Delta\cos\xi(5+3\cos^2\xi)}{3\sqrt3(3+\cos2\xi)^2}\hat{e}^7\wedge\hat{e}^8\bigg],\\
\hat{F}_3&=&8\Delta^{\ft32}3^{-\ft54}\bigg[\frac{8\sqrt2\sin\xi(2+9\cos^2\xi+5\cos^4\xi)}{(3+\cos 2\xi)\Xi}\hat{e}^4\wedge\hat{e}^5\wedge\hat{e}^9+\frac{(3+\cos 2\xi)^{\ft32}}{\sqrt{\Xi}}\hat{e}^4\wedge\hat{e}^6\wedge\hat{e}^7\nn\\
&&\qquad\qquad+\frac{4\cos\xi(4+3\cos^2\xi+\cos^4\xi)}{(3+\cos 2\xi)\sqrt{\Xi}}\hat{e}^4\wedge\hat{e}^8\wedge\hat{e}^9
+\frac{(3+\cos 2\xi)^{\ft32}}{\sqrt{\Xi}}\hat{e}^5\wedge\hat{e}^6\wedge\hat{e}^8\nn\\
&&\qquad\qquad -\frac{4\cos\xi(4+3\cos^2\xi+\cos^4\xi)}{(3+\cos 2\xi)\sqrt{\Xi}}\hat{e}^5\wedge\hat{e}^7\wedge\hat{e}^9+\frac{\sqrt2 \sin\xi}{3+\cos2\xi}\hat{e}^7\wedge\hat{e}^8\wedge\hat{e}^9\bigg],\\
\hat{F}_4&=&\ft{16}9\Delta^2\bigg[-\frac{2\cos\xi(4+3\cos^2\xi+\cos^4\xi)}{(3+\cos 2\xi)^{3/2}}\hat{e}^4\wedge\hat{e}^5\wedge\hat{e}^6\wedge\hat{e}^9+\frac{( 5+3 \cos2\xi)}{2} \hat{e}^4\wedge\hat{e}^5\wedge\hat{e}^7\wedge\hat{e}^8\nn\\
&&\qquad\quad +\frac{\sqrt{\Xi}\sin\xi}{\sqrt2(3+\cos 2\xi)^{3/2}}\hat{e}^4\wedge\hat{e}^6\wedge\hat{e}^8\wedge\hat{e}^9-\frac{\sqrt{\Xi}\sin\xi}{\sqrt2(3+\cos 2\xi)^{3/2}}\hat{e}^5\wedge\hat{e}^6\wedge\hat{e}^7\wedge\hat{e}^9\nn\\
&&\qquad\quad -\frac{2\cos\xi(4+9\cos^2\xi+3\cos^4\xi)}{(3+\cos 2\xi)^{3/2}}\hat{e}^6\wedge\hat{e}^7\wedge\hat{e}^8\wedge\hat{e}^9\bigg]+\frac{3\sqrt3}{8}{e}^0\wedge{e}^1\wedge{e}^2\wedge{e}^3.
\eea
Finally, the dilaton is given by
\be
e^{-\frac{3}{2}\hat{\phi}}=\frac{\Delta\Xi}{3\sqrt{3}(\cos 2 \xi +3)}.
\ee

\section*{Acknowledgements}
We are grateful to Hermann Nicolai, Chris Pope, Ergin Sezgin, Oscar Varela and Ning Su for discussions. The work of J.R. is partially supported by DOE grant DE-FG02-13ER42020. Y.P. is supported by the Alexander von Humboldt fellowship.

\end{document}